# Masses of Single, Double and Triple Heavy baryons in the Hyper-Central Quark Model by Using GF-AEIM

M. Abu-shady[1] and H. M. Fath-Allah[2]

Department of Mathematics and Computer Sciences, Faculty of Science, Menoufia University, Egypt[1]

Higher Institute of Engineering and Technology, Menoufia, Egypt[2]

## Abstract

In this paper, we calculate single, double and triple heavy baryons masses using hyper-central approach in the two cases. The first case, considering potential is a combination of Coulombic, linear confining and harmonic oscillator terms. The second case, we add the hyperfine interaction. The hyper-radial Schrödinger equation in the two cases is solved to obtain energy eigenvalues and the baryonic wave function by using the generalized fractional analytical iteration method (GF-AEIM). The present results are a good agreement with experimental data and are improved with other recent works.

**Keywords:** heavy baryons; hypercentral model; Generalized Fractional Exact solution.

## 1. Introduction

The constituent quark model (CQM), which is based on a hypercentral approach, has lately become popular for describing baryon internal structure [1–5]. Although the various theories are somewhat distinct, the heavy baryon spectrum is usually well described. Understanding the dynamics of QCD at the hadronic scale necessitates research into hadrons containing heavy quarks [6-10]. Due to the experimental observation of several heavy flavour baryons, heavy baryon characteristics have become a subject of considerable attention in recent years. All single charm quark-carrying spatial-ground-state baryons have been detected, and their masses have been calculated. Many spin-$\frac{1}{2}$ b-baryons $\Sigma b, \Xi b$, and $\Omega b$, as well as spin-$\frac{3}{2}$ baryons have been identified [11–15]. The doubly heavy baryons, which are made up of two heavy quarks and one light quark, are particularly intriguing because they offer a new platform for simultaneously exploring heavy quark symmetry and chiral dynamics [16-18]. There are a lot of theoretical models, the mass of the doubly heavy baryon $\Xi^{++}cc$ is predicted to be in the range 3.5~ 3.7

GeV. The mass splitting between $\Xi^{++}_{cc}$ and $\Xi^{+}_{cc}$ is predicted to be several MeV due to the mass difference of the light quarks u, d. The predicted mass in lattice QCD is about 3.6 GeV, which is quite close to the LHCb observation. The lifetimes of $\Xi^{++}_{cc}$ and $\Xi^{+}_{cc}$ are predicted to be quite long, 50∼ 250 and 200∼ 700 fs [19-28], respectively. In Refs. [29-32] calculated heavy flavor baryons containing single and double charm (beauty) quarks with light flavor combinations and considered the confinement potential as hypercentral Coulomb plus power potential with power index $v$. In Ref. [33], the author calculated baryons using Feynman–Hellmann theorem and semi-Fempirical mass formula within the framework of a non-relativistic constituent quark model. In Ref. [34], the author studied heavy-flavor baryons by using the Bethe-Salpeter equation in the heavy-quark limit and calculated the Isgur-Wise function. In Ref. [35], the author calculated different properties of single heavy-flavor baryons using heavy- quark symmetry in the non-relativistic quark model. In Ref. [36], the author investigated charmed baryons and spin-splittings in quenched lattice QCD. In Ref. [7], the author evaluated ground-state magnetic moments of heavy baryons in the relativistic quark model using heavy-hadron chiral perturbation theory. In Refs. [37-38], the authors solved the Schrodinger equation using iteration method to obtain masses of heavy baryons containing of single, double and triple in hyper-central approach with confining interaction and hyperfine interaction. In Ref. [39], the authors solved the Schrodinger equation using a variational method to obtain masses of single, double and triple in the hyper-central approach with confining interaction and hyperfine interaction. In Ref. [40], the authors use two potential the first potential is Cornell potential and the second potential is the same potential of Ref. [37-38] and they solved the Schrodinger equation numerically to obtain single, double and triple baryon masses. In Ref. [41], the authors obtained the masses of heavy-flavor baryon masses by using the non-relativistic quark model with hyper-central Coulomb plus linear potential and Coulomb plus harmonic oscillator potential. In Ref. [42], the authors obtained mass spectra of the doubly heavy baryons that the two heavy quarks inside a baryon form a compact heavy 'diquark core' in a color anti-triplet, and bind with the remaining light quark into a colorless baryon. In Ref. [43], the author calculated masses of the ground-state baryons consisting of three or two heavy and one light quarks in the framework of the relativistic quark model and masses of the triply and doubly heavy baryons are obtained by using the perturbation theory for the spin-independent and spin-dependent parts of the three-quark Hamiltonian. In Ref. [44], the author studied heavy baryons within Isgur-wise formalism by using the extended Cornell potential and solved the Schrodinger equation using iteration to obtain eigenvalues of energy and baryonic wave function.

In the present work, we employ generalized fractional iteration method and we calculate the masses of heavy-flavor baryons containing single, double and triple in the ground state in two cases. The first case, in the absence hyperfine interaction and the second case, in the presence hyperfine interaction.

This paper is arranged as follows. In Sec. 2, we display interaction potential. In Sec. 3, the theoretical method is explained. In Sec. 4, the results and discussion are written. In Sec. 5, the conclusion is written.

## 2. Interaction potential

In Ref. [37, 40], the considered potential is a combination of the Columbic-like term plus a linear confining term $(ax - c/x)$, as employed by QCD [45-46] and we have added the harmonic oscillator potential which has the form $ax^2$ as follows

$$V(x) = a x^2 + b x - \frac{c}{x}, \qquad (1)$$

where x is the relative quark pair coordinate. a, b and c are constants. We consider in the present potential hyperfine interaction potentials. In the second case, we have added hyperfine interaction potentials ($H_s(x)$, $H_I(x)$ and $H_{SI}(x)$). The nonperturbative confining interaction potential is the potential as defined in (1). The nonconfining potential due to the exchange interactions contains a δ-like term, an illegal operator term.[47]. We have modified it by a Gaussian of the quark pair relative distance the non-confining spin-spin interaction potential is proportional to a δ-function which is an illegal operator term. We modify it to a Gaussian function of the relative distance of the quark pair

$$H_s = A_S \frac{S_1 . S_2}{(\sqrt{\pi}\, \sigma_S)^3} Exp\left(\frac{-x^2}{\sigma_S^2}\right), \qquad (2)$$

where $s_i$ is the spin operator of the ith quark ($s_i = \sigma_i/2$), with $\sigma_i$ being the vector of Pauli matrices) and $A_S$ and $\sigma_S$ are constants. Other spin as well as isospin-dependent interaction potentials can arise from quark-exchange interactions. We conclude that two additional terms should be added to the Hamiltonian for quark pairs which result in hyperfine interactions similar to Eq. (3). The first one depends on isospin only and has the form [37, 47]

$$H_I = A_I \frac{t_1.t_2}{(\sqrt{\pi}\,\sigma_I)^3} Exp\left(\frac{-x^2}{\sigma_I^2}\right), \qquad (3)$$

where $t_i$ is the isospin operator of the ith quarks, and $A_I$ and $\sigma_I$ are constants. The second one is a spin-isospin interaction given by [37, 47]:

$$H_{SI} = A_{SI} \frac{(S_1.S_2)(t_1.t_2)}{(\sqrt{\pi}\sigma_{SI})^3} Exp\left(\frac{-x^2}{\sigma_{SI}^2}\right), \qquad (4)$$

where $s_i$ and $t_i$ are the spin and isospin operators of the ith quark, respectively, and $A_{SI}$ and $\sigma_{SI}$ are constants. Then, from Eqs. (2-4) the hyperfine interaction (a non-confining potential) is given by

$$H_{int} = H_s(x) + H_I(x) + H_{SI}(x), \qquad (5)$$

The parameters of the hyperfine interaction (5) are given in Table 1.

**Table (1).** Constituent hyperfine– potential parameters used in cases I and II [47-48]

| Parameter | Value |
|---|---|
| $A_S$ | 67.4 $(fm)^2$ |
| $\sigma_S$ | 4.76 $(fm)$ |
| $A_I$ | 51.7 $(fm)^2$ |
| $\sigma_I$ | 1.57 $(fm)$ |
| $A_{SI}$ | -106.2 $(fm)^2$ |
| $\sigma_{SI}$ | 2.31 $(fm)$ |

## 3. Theoretical method

### 3.1. Generalized fractional Derivative

Fractional derivative plays an important role in the applied science. Riemann-Liouville and Riesz and Caputo give a good formula that allows to apply boundary and initial conditions as in Ref. [49].

$$D_t^\alpha(r) = \int_{r_0}^r K_a(r-s)\, f^{(n)}(s)\, d(s), \qquad r > r_0 \qquad (6)$$

with

$$K_a(r-s) = \frac{(r-s)^{n-\alpha-1}}{\Gamma(n-\alpha)}, \tag{7}$$

where, $f^{(n)}$ is the n the derivative of the function f(t), and $K_a(r-s)$ is the kernel, which is fixed for a given real number α. The kernel $K_a(r-s)$ has singularity at r = s. Caputo and Fabrizio [50] suggested a new formula of the fractional derivative with smooth exponential kernel of the form to avoid the difficulties that found in Eq. (6)

$$D_t^\alpha = \frac{M(a)}{1-\alpha} \int_{r_0}^{r} \exp(\frac{\alpha(r-s)}{1-\alpha}) \, \dot{y}(s) \, d(s), \tag{8}$$

where M(a) is a normalization function with M(0) = M(1) =1.

A new formula of fractional derivative called generalized fractional derivative (CFD) is proposed [51].

$$D^\alpha [\, f_{nl}(r^\alpha)] = k \, r^{1-\alpha} \, \dot{f}_{nl}(r), \tag{9}$$

$$D^\alpha [\, D^\alpha f(r^\alpha)] = k^2 \, [(1-\alpha) \, r^{1-2\alpha} \, \dot{f}_{nl}(r) + r^{2-2\alpha} \, f_{nl}''(r)], \tag{10}$$

where, $k = \frac{\Gamma[\beta]}{\Gamma[\alpha-\beta+1]}$

with $0 < \alpha \leq 1, 0 < \beta \leq 1$.

### 3.2. Generalized Fractional Exact solution method of the Radial Schrödinger Equation for the Confining Potential

The baryon as bound state of three constituent quarks, we define the configuration of three particles by two the Jacobi coordinates ρ and λ as [29, 37, 38, 47, 55, 56,57]

$$\vec{\rho} = \frac{1}{\sqrt{2}} (\vec{r_1} - \vec{r_2}), \tag{11}$$

$$\vec{\lambda} = \frac{1}{\sqrt{6}} (\vec{r_1} + \vec{r_2} - 2\vec{r_3}), \tag{12}$$

where

$$m_\rho = \frac{2 m_1 m_2}{(m_1 + m_2)}; \qquad m_\rho = \frac{3 m_3 (m_1 + m_2)}{2(m_1 + m_2 + m_3)}$$

Here $m_1$, $m_2$ and $m_3$ are the constituent quark masses. Instead of $\rho$ and $\lambda$, one can introduce the hyperspherical coordinates, which are given by the angles $\Omega_\rho = (\theta_\rho, \varphi_\rho)$ and $\Omega_\lambda = (\theta_\lambda, \varphi_\lambda)$ together with the hyperradius x and the hyperangle, $\xi$ defined, respectively by [47]

$$x = \sqrt{\rho^2 + \lambda^2}, \qquad \xi = \tan^{-1}\left(\frac{\rho}{\lambda}\right). \tag{13}$$

Therefore, the Hamiltonian will be

$$H = \frac{p_\rho^2}{2 m_\rho} + \frac{p_\lambda^2}{2 m_\lambda} + V(\rho, \lambda) = \frac{p^2}{2 m} + V(x) \tag{14}$$

In the hypercentral constituent quark model (hCQM), the quark potential, V, is assumed to depend on the hyper radius x only, that is to be hypercentral. Therefore, $v = v(x)$ is in general a three-body potential, since the hyper radius x depends on the coordinates of all the three quarks. Since the potential depends on x only. In the three-quark wave function one can factor out the hyperangular part, which is given by hyperspherical harmonics. The remaining hyperradial part of the wave function is determined by hypercentral Schrödinger equation [48, 58]

$$\left[\frac{d^2}{dx^2} + \frac{5}{x}\frac{d}{dx} - \frac{\gamma(\gamma+4)}{x^2}\right] \Psi_{v,\gamma}(x) = -2 m (E - V(x)) \Psi_{v,\gamma}(x), \tag{15}$$

where $\Psi_{v,\gamma}(x)$ is the hyperradial wave function and $\gamma$ is the grand angular quantum number given by $\gamma = 2n + l_\rho + l_\lambda$; $l_\rho$ and $l_\lambda$ are the angular momenta associated with the $\rho$ and $\lambda$ variables and n is a non-negative integer number. v determines the number of the nodes of the wave function and m is the reduced mass [48]

$$m = \frac{2 m_\rho m_\lambda}{m_\rho + m_\lambda} \tag{16}$$

Now we want to solve the hyperradial Schrödinger equation for the three-body potential interaction (1). The wave function is factorized similarly to the central potential case. The transformation

$$\Psi_{v,\gamma}(x) = x^{-\frac{5}{2}} \phi_{v,\gamma}(x) \tag{17}$$

reduces Equ. (15) to the form

$$[\frac{d^2}{dx^2} + 2m(E - V(x)) - \frac{(2\gamma+3)(2\gamma+5)}{4x^2}]\phi_{v,\gamma}(x) = 0, \tag{18}$$

Assume $z = xA$, $x = \frac{z}{A}$

where $A = 1$ GeV

then, Eq. (1) becomes

$$V(z) = \frac{az^2}{A^2} + \frac{bz}{A} - \frac{cA}{z} \tag{19}$$

We note that Eq. (18) becomes

$$[\frac{d^2}{dz^2} + \frac{2m}{A^2}(E - V(z)) - \frac{(2\gamma+3)(2\gamma+5)}{4z^2}]\phi_{v,\gamma}(z) = 0, \tag{20}$$

The fractional of Eq. (20) by using Eq. (10),

$$D^\alpha[D^\alpha \phi_{v,\gamma}(z^\alpha)] = [-\frac{2m}{A^2}(E - V(z^\alpha)) + \frac{(2\gamma+3)(2\gamma+5)}{4z^2}]\phi_{v,\gamma}(z^\alpha) \tag{21}$$

where

$$V(z^\alpha) = \frac{az^{2\alpha}}{A^2} + \frac{bz^\alpha}{A} - \frac{cA}{z^\alpha} \tag{22}$$

and, we assume that,

$$\phi_{v,\gamma}(z^\alpha) = z^{-\left(\frac{1-\alpha}{2}\right)} R_{v,\gamma}(z) \tag{23}$$

By substituting Eqs. (9-10, 22, 23) into Equ. (21), we obtain the following equation

$$[\frac{d^2}{dz^2} + \varepsilon z^{2\alpha-2} - a_1 z^{4\alpha-2} - b_1 z^{3\alpha-2} + c_1 z^{\alpha-2} - \frac{(2\gamma+3)(2\gamma+5)}{4k^2 z^2} + \frac{\left(\frac{-\alpha^2}{4}+\frac{1}{4}\right)}{z^2}] R_{v,\gamma}(z^\alpha) = 0 \tag{24}$$

where,

$$\varepsilon = \frac{2mE}{A^2 k^2}, \quad a_1 = \frac{2ma}{A^4 k^2}, \quad b_1 = \frac{2mb}{A^3 k^2}, \quad c_1 = \frac{2mc}{A k^2}. \tag{25}$$

The analytical exact iteration method (AEIM) requires making the following ansatz [37] as follows

$$R_{v,\gamma}(z^\alpha) = f(z^\alpha) \exp[g(z^\alpha)], \tag{26}$$

where

$$f_{(z^\alpha)}=\begin{cases} 1, & n = 0, \\ \prod_{i=1}^{n}(z^\alpha - \alpha_i^{(n)}) & n = 1,2,\ldots \end{cases} \quad (27)$$

$$g(z^\alpha) = -\frac{1}{2}\alpha_1 z^{2\alpha} - \beta_1 z^\alpha + \delta_1 \ln z^\alpha, \quad \alpha_1 > 0, \beta_1 > 0 \quad (28)$$

It is clear that $f(z)$ are equivalent to the Laguere polynomials at $\alpha = 1$. From Eq. (21), we obtain

$$R_{v,\gamma}''(z^\alpha) = [\, g_1''(z^\alpha) + g_1'^2(z^\alpha) + \frac{f''(z^\alpha) + 2f'(z^\alpha)g'(z^\alpha)}{f(z^\alpha)} \,] R_{v,\gamma}(z^\alpha). \quad (29)$$

$$a_1 z^{4\alpha-2} + b_1 z^{3\alpha-2} - \varepsilon z^{2\alpha-2} - c_1 z^{\alpha-2} + (\frac{(2\gamma+3)(2\gamma+5)}{4k^2} + \frac{\alpha^2}{4} - \frac{1}{4})z^{-2} = \alpha_1^2 \alpha^2 z^{4\alpha-2} + 2\alpha_1 \alpha^2 \beta_1 z^{3\alpha-2} + (-\alpha(2\alpha-1)\alpha_1 - 2\alpha_1 \alpha^2 \delta + \beta_1^2 \alpha^2) z^{2\alpha-2} + (-\beta_1 \alpha(\alpha-1) - 2\beta_1 \alpha^2 \delta) z^{\alpha-2} + (-\delta\alpha + \delta^2 \alpha^2) z^{-2} \quad (30)$$

Now, comparing the coefficient of $z$ both sides of Eq. (30)

$$\alpha_1 = \frac{\sqrt{a_1}}{\alpha}, \quad (31)$$

$$\beta_1 = \frac{b_1}{2\alpha\sqrt{a_1}}, \quad (32)$$

$$c_1 = \beta_1 \alpha(\alpha-1) + 2\beta_1 \alpha^2 \delta \quad (33)$$

$$\varepsilon = \alpha(2\alpha-1)\alpha_1 + 2\alpha_1 \alpha^2 \delta - \beta_1^2 \alpha^2 \quad (34)$$

$$\delta = \frac{1}{2\alpha}[1 \pm \alpha\sqrt{\alpha^2 + \frac{1}{k^2}(4\gamma^2 + 16\gamma + 15)}\,] \quad (35)$$

Let us assume $\omega^2 = \frac{3k}{m}$ then $\omega = \sqrt{\frac{2a}{m}}$ as in Ref. [37]

The Eqs. (31, 32, 33) become,

$$\alpha_1 = \frac{m\omega}{A^2 \alpha k} \quad (36)$$

$$\beta_1 = \frac{b}{A\alpha\omega k} \quad (37)$$

$$c = \frac{bk((\alpha-1)+2\alpha\delta)}{2m\omega} \quad (38)$$

The energy eigenvalue for the mode $\nu = 0$ and grand angular momentum $\gamma$ from Eqs. (24, 31, 32, 35, 36, 37)

$$E_{0,\gamma} = \frac{\omega}{2} k (2\alpha - 1) + \omega k \alpha \delta - \frac{b^2}{2 m \omega^2}. \tag{39}$$

then from Eqs. (23), (26) and (35-36, 37) the normalized eigenfunctions are given as

$$\Psi_{0,\gamma} = N_{0,\gamma}\, x^{\frac{\alpha}{2} + \delta\, \alpha - 3} \operatorname{Exp}\left(\frac{-m\omega}{2\,\alpha\,A^2\,k} x^{2\alpha} - \frac{2mc}{k^2 A\,(\alpha(\alpha-1) + 2\alpha^2 \delta)} x^{\alpha}\right) \tag{40}$$

## 4. Results and Discussion

We calculate the baryon masses are given by three quark masses and the energy $E_{\nu\gamma}$ which is a function of a, b and $m_q$ in two cases, the first case without the hyperfine interaction masses and the second case, with the hyperfine interaction potential $\langle H_{int} \rangle$ treated as a perturbation. The first order energy correction from the nonconfining potential $\langle H_{int} \rangle$ can be obtained by using the unperturbed wave function [37].

### 4.1 The interaction potential without hyperfine interaction

In the first case, the Baryon mass then becomes the sum of quarks mass and energy, thus [60]

$$M = m_{q1} + m_{q2} + m_{q3} + E_{\nu\gamma} \tag{41}$$

**Table (2)**. The values of the used quark masses in two cases in GeV [59]

| $m_u$ | $m_d$ | $m_s$ | $m_c$ | $m_b$ |
|---|---|---|---|---|
| 0.330 | 0.335 | 0.310 | 1.6 | 4.980 |

We use the same potential of Refs. [37, 39, 40] but we solve the Schrodinger equation by using the generalized fractional analytical iteration method. The quark masses and potential parameters are listed in the Tables (1, 2) and the constants a, b and c of the potential and $\omega$ as in Ref. [40]. We note that the generalized fractional analytical iteration method plays an important role. In Table (3), we calculate single charm baryon masses in the ground state (masses are in GeV) at ($\alpha = \beta = 0.665$). The present results close with experimental data such that $\Sigma_c^+(udc)$ and are good compared with other works such that Ref. [37] total error is 0.19%. In Ref. [39], the total error is 5.3%. In Ref. [30], the total error is 0.53%. In Ref. [40], the total error is 0.26% and in Ref. [29], the total error is 0.46% but the total error of present work is 0.13%. In Table (4), we calculate single beauty baryon masses in the ground state (masses are in GeV) at ($\alpha$ = 0.56). The present results are close with experimental data and are a good agreement compared with other works such that Ref. [37] total error is 0.42%. In Ref. [40], the total error is 1.7%. In Ref. [5], the total error is 1.0275%. In Ref. [38], the total error is 0.26% and in Ref. [29], the total error is 0.9% but the total error of present work is 0.0375%. In Table (5), we calculate double charm and beauty baryon masses in ground state at ($\alpha$ = 0.1), we note that, the present results are a good agreement with recent works such that [30, 37, 40, 41, 42]. In Table (6), we calculate charm and beauty baryon masses in the ground state (masses are in GeV) at ($\alpha = \beta = 0.2$), our results are a good agreement with recent works such that [40, 42, 43].

**Table (3).** Single charm baryon masses in ground state (masses are in GeV) at ($\alpha = \beta = 0.665$). The last column shows the relative error in comparison to experimental data.

| Baryon | Present work | Exp. | Ref.[37] | Ref.[39] | Ref.[30] | Ref.[40] | Ref.[29] | Relative Error |
|---|---|---|---|---|---|---|---|---|
| $\Sigma_c^{++}(uuc)$ | 2.448 | 2.454 | 2.452 | 2.318 | 2.443 | 2.459 | 2.425 | 0.2% |
| $\Sigma_c^{+}(udc)$ | 2.453 | 2.453 | 2.457 | 2.323 | 2.460 | 2.461 | - | 0.0% |
| $\Sigma_c^{0}(ddc)$ | 2.458 | 2.454 | 2.461 | 2.328 | 2.477 | 2.462 | 2.460 | 0.2% |
| Total error | 0.13% | - | 0.19% | 5.3% | 0.53% | 0.26% | 0.46% | |

**Table (4).** Single beauty baryon masses in ground state (masses are in GeV) at ($\alpha = \beta = 0.56$). The last column shows the relative error in comparison to experimental data.

| Baryon | P.W | Exp | Ref.[29] | Ref.[37] | Ref.[38] | Ref.[40] | Relative error |
|---|---|---|---|---|---|---|---|
| $\Sigma_b^+(uub)$ | 5.806 | 5.807 | 5.772 | 5.807 | 5.816 | 5.834 | 0.02% |
| $\Sigma_b^-(ddb)$ | 5.817 | 5.815 | 5.816 | 5.818 | 5.821 | 5.844 | 0.03% |
| $\Xi_b^0(usb)$ | 5.784 | 5.787 | 5.880 | 5.821 | 5.886 | 5.956 | 0.05% |
| $\Xi_b^-(dsb)$ | 5.789 | 5.792 | 5.903 | 5.826 | 5.887 | 5.961 | 0.05% |
| Total error | 0.0375% | - | 1.0275% | 0.42% | 0.9% | 1.7% | |

**Table (5).** Double charm and beauty baryon masses in ground state (masses are in GeV) at ($\alpha = \beta = 0.1$).

| Baryon | P.W | Ref.[30] | Ref.[37] | Ref.[40] | Ref.[41] | Ref.[42] |
|---|---|---|---|---|---|---|
| $\Xi_{cc}^{++}(ucc)$ | 3.622 | 3.730 | 3.583 | 3.703 | 3.676 | 3.601 |
| $\Xi_c^+(dcc)$ | 3.627 | 3.755 | 3.588 | 3.708 | 3.676 | - |
| $\Omega_{cc}^+(scc)$ | 3.600 | 3.857 | 3.592 | 3.846 | 3.815 | 3.592 |
| $\Xi_{bb}^0(ubb)$ | 10.395 | - | 10.284 | 10.467 | 10.340 | 10.182 |
| $\Omega_{bb}^-(sbb)$ | 10.373 | - | 10.239 | 10.606 | 10.454 | 10.276 |

**Table (6).** charm and beauty baryon masses in ground state (masses are in GeV) at ($\alpha = \beta = 0.2$).

| Baryon | P.W | Ref.[40] | Ref.[40] | Ref.[42] | Ref.[43] |
|---|---|---|---|---|---|
| $\Omega_{cb}^+(ucb)$ | 7.027 | 7.087 | 6.988 | 6.931 | 6.792 |
| $\Omega_{cb}^0(scb)$ | 7.01 | 7.226 | 7.103 | 7.033 | 6.999 |
| $\Omega_{ccb}^+(ccb)$ | 8.321 | 8.357 | 8.190 | - | 8.018 |
| $\Omega_{cbb}^0(cbb)$ | 11.706 | 11.737 | 11.542 | - | 11.280 |

## 4.2 The interaction potential with hyperfine interaction

In the second case, the Baryon mass then becomes the sum of quarks mass and energy with the hyperfine interaction potential <$H_{int}$> treated as a perturbation, thus as in Refs. [37, 38, 40]

$$<H_{int}> = \int \Psi H_{int} \Psi \, dx \tag{42}$$

$$M = m_{q1} + m_{q2} + m_{q3} + E_{v\gamma} + <H_{int}> \tag{43}$$

In this case, we also get a good results with experiment and theoretical works as in Table (7), we calculate single charm and beauty baryon masses in the ground state (masses are in GeV) at ($\alpha = 0.678$) in case charm and ($\alpha = 0.54$) in case beauty. The present results is a good agreement with experimental data such as $\Sigma_c^{++}(uuc)$ and are good compared with other works such that Ref. [37] total error is 0.953%. In Ref. [39], the total error is 2.765% but the total error of the present work is 0.655%. In Table (8), we calculate double and triple charm and beauty baryon masses in ground state (masses are in GeV) at ($\alpha = 0.39$). The present result is a good with recent works such that Refs. [37, 38, 40, 42, 43]. In Table (9), we calculate charm and beauty baryon masses in the ground state (masses are in GeV) at ($\alpha = 0.2$). The present result is a good agreement with recent works such that Refs. [37, 40, 42, 43].

In Refs. [37-38], the authors solved the Schrodinger equation using iteration method and the considered potential is a combination of Coulombic, linear confining and harmonic oscillator terms to obtain masses of heavy baryons containing of single, double and triple in the hyper-central approach with confining interaction and hyperfine interaction in the first case, total error in Ref. [37] is 0.19% and 0.42% when they calculated single charm and beauty baryon masses in the ground state respectively, in the second case, the total error is 0.953% when they calculated single charm and beauty baryon masses in ground state. In Ref. [38], the total error is 0.9% when they calculated single beauty baryon masses in the ground state. In Ref. [39], the authors solved the Schrodinger equation using a variational method and the considered potential is Coulomb as well as linear confining terms and the spin–isospin dependent potential to obtain masses of single, double and triple in the hyper-central approach with confining interaction and hyperfine interaction, in the first case total error is 5.3% when they calculated single charm baryon masses in ground state and the second case total error is 2.765%% when calculated single charm and beauty baryon masses in ground state. In Ref. [30], the authors obtained the masses of the baryons containing single charm and beauty quark in the presence confinement potential is assumed in the hyper central co-ordinates of the coulomb plus power potential form, in the first case, total error is 0.53% when they calculated Single charm baryon masses in ground state. In Ref. [40], the authors use two potential the second potential is the same potential and the hyper-central approach of Ref. [37-38] and use the Cornell potential and the hyper-central approach but they solved the Schrodinger equation numerically to obtain single, double and triple baryon masses and in the first case total error is 0.26% and 1.7% when they calculated single charm and beauty baryon masses in ground state. In Ref. [29], in the first case, total error is 0.46% and1.0275% when calculated single charm and beauty baryon masses in the ground state. The authors obtained the masses of heavy-flavor baryon masses by using the non-relativistic quark model with hyper-central Coulomb plus linear potential and Coulomb plus harmonic oscillator potential in Ref. [41]. In Ref. [42], the authors obtain mass spectra of the doubly heavy baryons are computed assuming that the two heavy quarks inside a baryon form a compact heavy 'diquark core' in a color anti-triplet, and bind with the remaining light quark into a colorless baryon. The two reduced two-body problems are described by the relativistic Bethe-Salpeter equations (BSEs) with the relevant QCD inspired kernels. In Ref. [43], the author calculates the masses of the ground-state baryons consisting of three or two heavy and one light quark in the framework of the relativistic quark model and the hyperspherical expansion. The masses of the triply and doubly heavy baryons are obtained by using the perturbation theory for the spin-independent and spin-dependent parts of the three-quark Hamiltonian.

**Table. (7)** Single charm and beauty baryon masses in ground state (masses are in GeV) at ($\alpha = \beta = 0.678$) and ($\alpha = \beta = 0.54$). The last column shows the relative error in comparison to experimental data.

| Baryon | $I(j^p)$ | $<H_{int}>$ | P.W | Exp | Ref.[37] | Ref.[39] | Relative error |
|---|---|---|---|---|---|---|---|
| $\Sigma_c^{++}(uuc)$ | $1(\frac{1}{2}^+)$ | 0.00292501 | 2.454 | 2.454 | 2.452 | 2.318 | 0.0% |
| | $1(\frac{3}{2}^+)$ | 0.0157481 | 2.467 | 2.518 | 2.581 | 2.446 | 2% |
| $\Sigma_c^+(udc)$ | $1(\frac{1}{2}^+)$ | 0.00292876 | 2.460 | 2.453 | 2.457 | 2.323 | 0.3% |
| | $1(\frac{3}{2}^+)$ | 0.0157452 | 2.472 | 2.518 | 2.586 | 2.451 | 1.8% |
| $\Sigma_c^0(ddc)$ | $1(\frac{1}{2}^+)$ | 0.00293254 | 2.465 | 2.454 | 2.461 | 2.328 | 0.4% |
| | $1(\frac{1}{2}^+)$ | 0.0157422 | 2.478 | 2.518 | 2.591 | 2.456 | 1.6% |
| $\Sigma_b^+(uub)$ | $1(\frac{1}{2}^+)$ | 0.00636099 | 5.807 | 5.807 | 5.807 | 5.700 | 0.0% |
| | $1(\frac{3}{2}^+)$ | 0.0135397 | 5.815 | 5.829 | 5.936 | 5.826 | 0.2% |
| $\Sigma_b^-(ddb)$ | $1(\frac{1}{2}^+)$ | 0.0063621 | 5.818 | 5.815 | 5.818 | 5.708 | 0.05% |
| | $1(\frac{3}{2}^+)$ | 0.013539 | 5.826 | 5.836 | 5.946 | 5.836 | 0.2% |
| Total error | - | - | 0.655% | - | 0.953% | 2.765% | |

**Table (8).** Double and triple charm and beauty baryon masses in ground state (masses are in GeV) at ($\alpha = \beta = 0.39$).

| Baryon | $I(j^p)$ | P.W | Ref.[37] | Ref.[40] | Ref.[40] | Ref.[43] | Ref.[39] | Ref.[42] |
|---|---|---|---|---|---|---|---|---|
| $\Xi_{cc}^{++}(ucc)$ | $\frac{1}{2}(\frac{1}{2}^+)$ | 3.608 | 3.583 | 3.703 | 3.532 | 3.510 | 3.597 | 3.601 |
| | $\frac{1}{2}(\frac{3}{2}^+)$ | 3.760 | 3.722 | 3.765 | 3.623 | 3.548 | 3.708 | 3.703 |
| $\Omega_{cc}^+(scc)$ | $0(\frac{1}{2}^+)$ | 3.586 | 3.592 | 3.846 | 3.667 | 3.719 | 3.718 | 3.710 |
| | $0(\frac{3}{2}^+)$ | 3.738 | 3.731 | 3.904 | 3.758 | 3.746 | 3.847 | 3.814 |

| Baryon | $I(J^P)$ | | | | | | |
|---|---|---|---|---|---|---|---|
| $\Xi_c^+(dcc)$ | $\frac{1}{2}(\frac{1}{2}^+)$ | 3.613 | 3.588 | 3.708 | 3.537 | 3.510 | 3.584 | 3.606 |
| | $\frac{1}{2}(\frac{3}{2}^+)$ | 3.765 | 3.726 | 3.770 | 3.629 | 3.548 | 3.713 | 3.706 |
| $\Omega_{ccc}^{++}(ccc)$ | $0(\frac{3}{2}^+)$ | 5.053 | 4.842 | 5.035 | 4.880 | 4.803 | 4.978 | - |
| $\Xi_{bb}^0(ubb)$ | $\frac{1}{2}(\frac{1}{2}^+)$ | 10.380 | 10.284 | 10.467 | 10.334 | 10.130 | 10.339 | 10.182 |
| | $\frac{1}{2}(\frac{3}{2}^+)$ | 10.532 | 10.427 | 10.525 | 10.431 | 10.144 | 10.468 | 10.214 |
| $\Omega_{bb}^0(dbb)$ | $\frac{1}{2}(\frac{1}{2}^+)$ | 10.386 | 10.289 | - | - | 10.130 | 10.344 | - |
| | $\frac{1}{2}(\frac{3}{2}^+)$ | 10.538 | 10.432 | - | - | 10.144 | 10.473 | - |
| $\Omega_{bb}^-(sbb)$ | $0(\frac{1}{2}^+)$ | 10.359 | 10.293 | 10.606 | 10.397 | 10.424 | 10.478 | 10.276 |
| | $0(\frac{3}{2}^+)$ | 10.510 | 10.436 | 10.664 | 10.495 | 10.432 | 10.607 | 10.309 |
| $\Omega_{bbb}^-(bbb)$ | $0(\frac{3}{2}^+)$ | 15.208 | 14.810 | 15.175 | 15.023 | 14.569 | 15.118 | - |

**Table (9)** charm and beauty baryon masses in ground state (masses are in GeV) at ($\alpha\,\beta = 0.2$).

| Baryon | $I(J^P)$ | P.W | Ref. [37] | Ref.[40] | Ref.[42] | Ref.[43] |
|---|---|---|---|---|---|---|
| $\Omega_{cb}^+(ucb)$ | $\frac{1}{2}(\frac{1}{2}^+)$ | 6.951 | 6.935 | 7.078 | 6.931 | 6.792 |
| | $\frac{1}{2}(\frac{3}{2}^+)$ | 7.103 | 7.076 | 7.145 | 6.997 | 6.827 |
| $\Omega_{cb}^0(scb)$ | $\frac{1}{2}(\frac{1}{2}^+)$ | 6.929 | 6.945 | 7.226 | 7.033 | 6.999 |
| | $\frac{1}{2}(\frac{3}{2}^+)$ | 7.081 | 7.085 | 7.284 | 7.101 | 7.024 |
| $\Omega_{ccb}^+(ccb)$ | $\frac{1}{2}(\frac{1}{2}^+)$ | 8.244 | 8.038 | 8.357 | - | 8.018 |
| | $\frac{1}{2}(\frac{3}{2}^+)$ | 8.397 | 8.186 | 8.415 | - | 8.025 |
| $\Omega_{cbb}^0(cbb)$ | $0(\frac{1}{2}^+)$ | 11.631 | 11.363 | 11.737 | - | 11.280 |
| | $0(\frac{3}{2}^+)$ | 11.782 | 11.512 | 11.795 | - | 11.287 |

## 5. Conclusion

. In this paper, we employ generalized fractional iteration method and calculate the masses of heavy-flavor baryons containing single, double and triple in the ground state in two cases. The first case, in the absence hyperfine interaction and the second case, in the presence hyperfine interaction. We calculate the three-body analytical solution of the hyper-central Schrodinger equation using generalized fractional analytical iteration method. This method plays an important role in two cases because in two cases and we obtain the results are close with experimental data and are good compared with other works.

In the case of the interaction potential without hyperfine interaction, we calculate single charm baryon masses are close with experimental data and are a good agreement compared with other works because in Ref. [37], the total error is 0.19%, in Ref. [39], the total error is 5.3%, in Ref. [30], the total error is 0.53%, in Ref. [40], the total error is 0.26% and in Ref. [29], the total error is 0.46% but the total error of present work is 0.13%. When we calculate single beauty baryon mass such that $\Sigma_c^{++}(uuc)$ is close experimental data and are a good agreement compared with other works such that Ref. [37] total error is 0.42%, in Ref. [40] total error is 1.7% , in Ref. [29] total error is 1.0275%, in Ref. [38] total error is 0.26% and in Ref. [38] total error is 0.9% but the total error of the present work is 0.0375%.

In the case of the interaction potential with hyperfine interaction, we calculate single charm and beauty baryon masses are close with experimental data such that $\Sigma_c^{++}(uuc)$ and are a good agreement compared with other works because in Ref. [37], the total error is 0.953%. In Ref. [39], the total error is 2.765% but the total error of the present work is 0.655%. We conclude that in the case the interaction potential without and with hyperfine interaction in framework of GF-AEIM gives a good description of the heavy-flavor baryons in comparison with experimental data and other works.